\documentclass[twocolumn]{aastex631}

\usepackage{tabularx}
\begin{document}

\title{JWST constraints on the UV luminosity density at cosmic dawn:
implications for 21-cm cosmology}

\author[0000-0002-1050-7572]{Sultan Hassan}
\altaffiliation{E-mail: sultan.hassan@nyu.edu}\altaffiliation{NHFP Hubble Fellow.}

\affiliation{Center for Cosmology and Particle Physics, Department of Physics, New York University, 726 Broadway, New York, NY 10003, USA}
\affiliation{Center for Computational Astrophysics, Flatiron Institute, 162 5th Ave, New York, NY 10010, USA}\affiliation{Department of Physics \& Astronomy, University of the Western Cape, Cape Town 7535,
South Africa}

\author[0000-0001-7964-5933]{Christopher C. Lovell}
\affiliation{Institute of Cosmology and Gravitation, University of Portsmouth, Burnaby Road, Portsmouth, PO1 3FX, UK}
\affiliation{Astronomy Centre, University of Sussex, Falmer, Brighton BN1 9QH, UK}

\author[0000-0002-6336-3293]{Piero Madau}
\affiliation{Department of Astronomy \& Astrophysics, University of California, Santa Cruz, 95064, CA, USA}
\affiliation{Dipartimento di Fisica ``G. Occhialini", Università degli Studi di Milano-Bicocca, Piazza della Scienza 3, I-20126 Milano, Italy}

\author[0000-0002-1416-8483]{Marc Huertas-Company}
\affil{Instituto de Astrof\'isica de Canarias, La Laguna, Tenerife, Spain}
\affil{Universidad de la Laguna, La Laguna, Tenerife, Spain}
\affil{Universit\'e Paris-Cit\'e, LERMA - Observatoire de Paris, PSL, Paris, France}
\author[0000-0002-6748-6821]{Rachel S. Somerville}
\affiliation{Center for Computational Astrophysics, Flatiron Institute, 162 5th Ave, New York, NY 10010, USA}
\author[0000-0001-5817-5944]{Blakesley Burkhart}
\affiliation{Department of Physics and Astronomy, Rutgers University,  136 Frelinghuysen Rd, Piscataway, NJ 08854, USA}
\affiliation{Center for Computational Astrophysics, Flatiron Institute, 162 5th Ave, New York, NY 10010, USA}
\author{Keri L. Dixon}
\affiliation{New York University Abu Dhabi, PO Box 129188, Abu Dhabi, United Arab Emirates}
\affiliation{Center for Astro, Particle and Planetary Physics (CAP$^3$), New York University Abu Dhabi}
\author[0000-0002-1109-1919]{Robert Feldmann}
\affiliation{Institute for Computational Science, University of Zurich, Zurich CH-8057, Switzerland}
\author[0000-0003-2539-8206]{Tjitske K. Starkenburg}
\affiliation{Department of Physics \& Astronomy and CIERA, Northwestern University, 1800 Sherman Ave, Evanston, IL 60201, USA}
\author[0000-0002-5077-881X]{John F. Wu}
\affiliation{Space Telescope Science Institute, 3700 San Martin Dr, Baltimore, MD 21218}
\affiliation{Department of Physics \& Astronomy, Johns Hopkins University, 3400 N Charles St, Baltimore, MD 21218}

\author[0000-0002-8896-6496]{Christian Kragh Jespersen}
\affiliation{Department of Astrophysical Sciences, Princeton University, Princeton, NJ 08544, USA}
\author[0000-0003-4679-1058]{Joseph D. Gelfand}
\affiliation{New York University Abu Dhabi, PO Box 129188, Abu Dhabi, United Arab Emirates}
\affiliation{Center for Astro, Particle and Planetary Physics (CAP$^3$), New York University Abu Dhabi}
\author[0000-0001-7072-570X]{Ankita Bera}
\affiliation{School of Astrophysics, Presidency University, 86/1 College Street, Kolkata 700073, India}
\affiliation{Department of Astronomy, University of Maryland, College Park, MD 20742, USA}



\begin{abstract}
An unprecedented array of new observational capabilities are starting to yield key constraints on models of the epoch of first light in the Universe. In this Letter we discuss the implications of the UV radiation background at cosmic dawn inferred by recent \textrm{JWST} observations for radio experiments aimed at detecting the redshifted 21-cm hyperfine transition of diffuse neutral hydrogen. Under the basic assumption that the 21-cm signal is activated by the Ly$\alpha$ photon field produced by metal-poor stellar systems, we show that a detection at the low frequencies of the \textrm{EDGES} and SARAS3 experiments may be expected from a simple extrapolation of the declining UV luminosity density inferred at $z\lesssim 14$ from \textrm{JWST} early galaxy data. Accounting for an early radiation excess above the CMB suggests a shallower or flat evolution to simultaneously reproduce low and high-$z$ current UV luminosity density constraints, which cannot be entirely ruled out, given the large uncertainties from cosmic variance and the faint-end slope of the galaxy luminosity function at cosmic dawn.
Our findings raise the intriguing possibility that a high star formation efficiency at early times may trigger the onset of intense Ly$\alpha$ emission at redshift $z\lesssim 20$ and produce a cosmic 21-cm absorption signal 200 Myr after the Big Bang.  
\end{abstract}



\section{Introduction} \label{sec:intro}

A number of observational facilities are currently or will soon probe the epoch of cosmic dawn ($z > 10$), when the first stars and galaxies are expected to have formed. Results from these facilities are expected to place important constraints on the first astrophysical sources of radiation, including their number density, ionising emissivity, as well as the physics of their formation.

The James Webb Space Telescope (\textrm{JWST}) is the current flagship space-based infrared observatory, and was specifically designed to probe the epoch of first light  as a one of the main scientific goals~\citep{robertson_galaxy_2022}. One of the first tantalising results from early-release \textrm{JWST} data has been the discovery of very high redshift candidate galaxies in NIRCam imaging \citep[\textit{e.g.}][]{adams_discovery_2022,atek_revealing_2023,bradley_high-redshift_2022,castellano_early_2022,castellano_high_2022,donnan_evolution_2022,donnan_abundance_2023,finkelstein_long_2022,finkelstein_ceers_2023,harikane22,morishita_physical_2023,naidu_two_2022,perez-gonzalez_life_2023,yan_first_2023}.
Not only are these galaxies at much higher redshifts than any galaxy discovered previously by the Hubble Space Telescope (\textrm{HST}), but they are also surprisingly bright.
Almost all galaxy formation models struggle to reproduce the number densities of these bright early systems \citep{finkelstein_ceers_2023}.
Additionally, after performing SED fitting on the measured fluxes, many authors obtain high stellar masses \citep{donnan_evolution_2022,harikane_comprehensive_2022,labbe_very_2022} that may be in tension with the astrophysics of early galaxy formation
\citep{boylan-kolchin_stress_2022,lovell_extreme_2023} (but see 
\citealt{2023arXiv230413755M,2023arXiv230411911P,2023arXiv230404348Y} for 
a different interpretation).
Given these challenges, it may be important to seek out independent measurements of the source population at very early epochs.
\begin{figure*}
    \centering
    \includegraphics[scale=0.4]{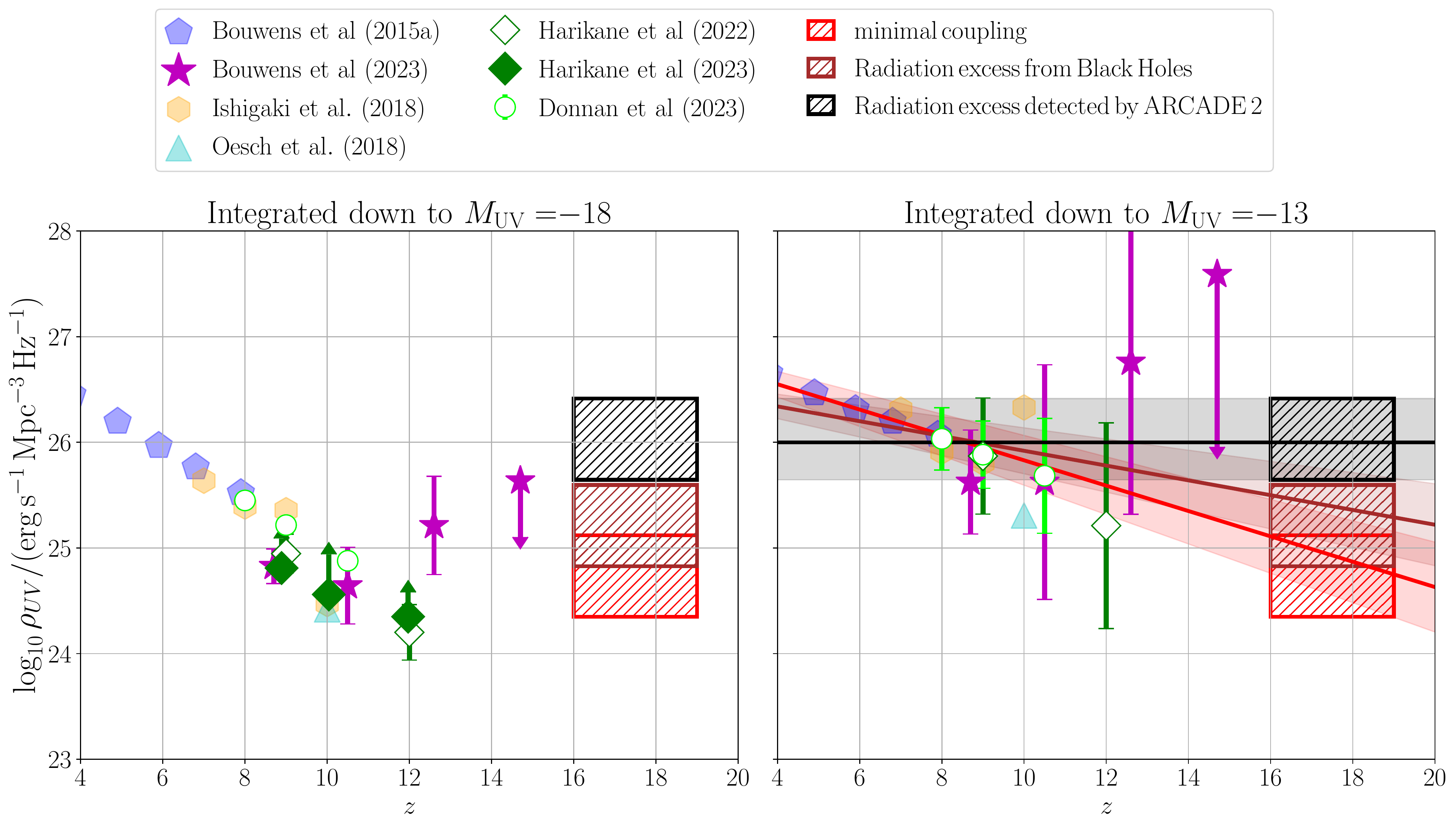}
    \caption{The galaxy UV luminosity density, $\rho_{\rm UV}$, from \textrm{HST} (\citealt[faint cyan traingles]{oesch18}, \citealt[faint orange hexagons]{ishigaki18}, \citealt[faint blue pentagons]{bouwens15}) and \textrm{JWST} (\citealt[open lime circles]{donnan23},~\citealt[open green diamonds]{harikane22},~\citealt[magenta stars]{bouwens2022}), using the measured LF from $z=4$ to $z=14$ for different magnitude cutoff $M_{\rm UV}=-18$ (left panel), and $M_{\rm UV}=-13$ (right panel). Error bars are added only to JWST data assuming 5\% and 10\% uncertainties in the faint-end slope at $z<10$ and $z>10$, respectively. The recently confirmed spectroscopic measurements reported by~\citet{2023arXiv230406658H} are shown as filled green diamonds. The red box depicts
    the UV luminosity density needed to produce a 21-cm global signal at $16 < z < 19$ 
    in the ‘minimal coupling’ regime \citep{madau18}. The brown and black boxes show the enhanced $\rho_{\rm UV}$ required by the presence of a radio background excess produced by early black holes~\citep{Ewall_2018} and following the claimed detection by ARCADE 2~\citep{Feng_2018}, respectively.  If \textrm{JWST} LF measurements could be extrapolated down to $M_{\rm UV}=-13$, the ensuing luminosity density would match the gradual redshift evolution predicted by \citet{madau18} and the updated fit (red line), providing enough 
    Ly$\alpha$ background radiation to mix the hyperfine levels of neutral hydrogen 200 Myr after the Big Bang. Due to the large uncertainty associated with cosmic variance/faint-end slope of the LF at these early epochs, the enhanced UV luminosity density required by the presence of a radio background excess (brown and black boxes) is also broadly consistent with current JWST and HST data, following shallow (brown line) or flat (black line) evolution in $\rho_{\rm UV}$.
    }
    \label{fig:rho}
\end{figure*}

\begin{figure}
    \includegraphics[scale=0.4]{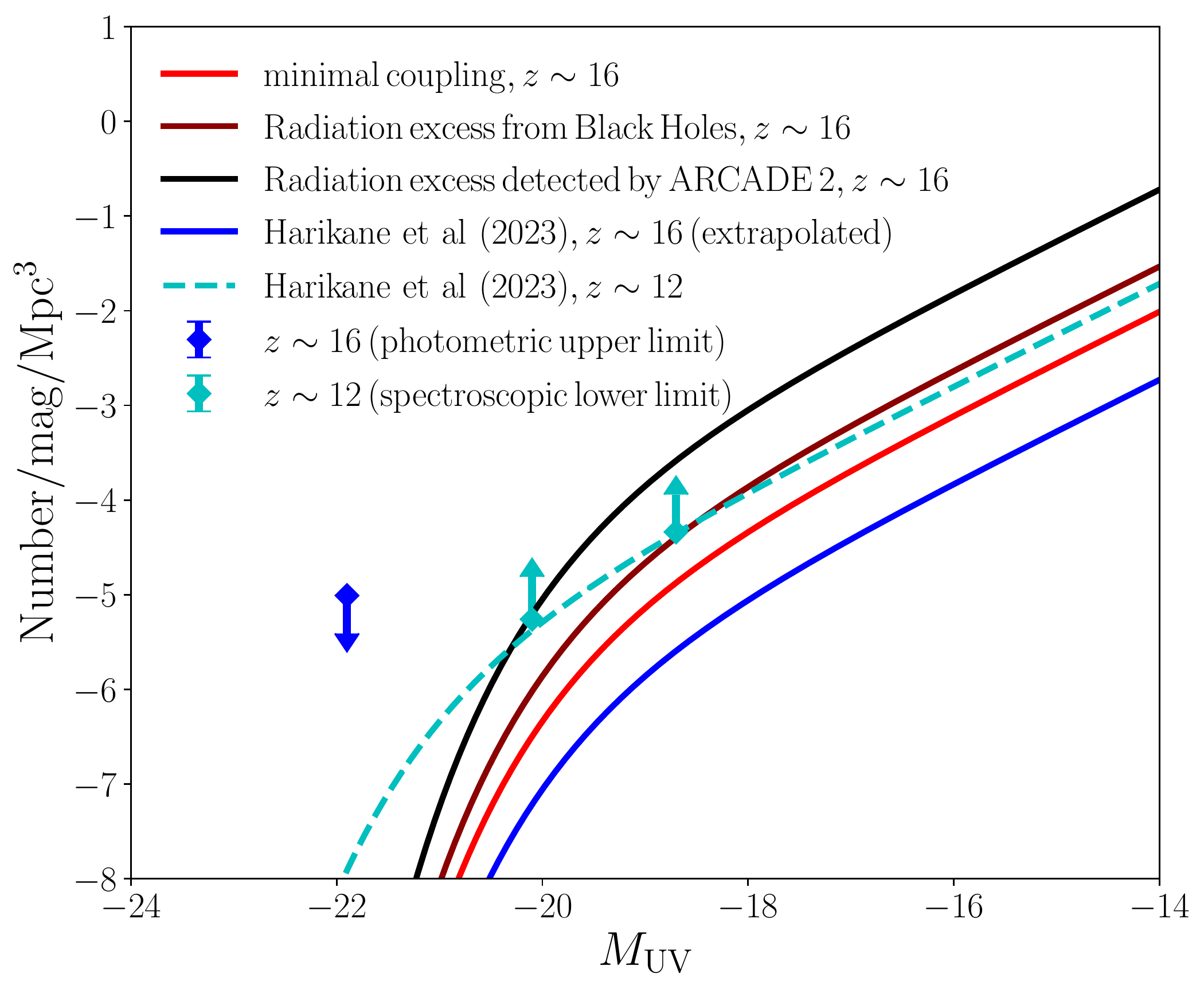}
\caption{Predicted galaxy UV luminosity function at $z=16$. The Schechter function parameter $\phi^{\star}$ has been normalized to yield the luminosity density required in
the minimal 21-cm coupling \citep[solid red line]{madau18}, the early black hole radio \citep[brown solid line]{Ewall_2018}, and the ARCADE 2 radio excess \citep[black solid line]{Feng_2018} scenarios, at fixed $M^{\star}_{\rm UV}=-21.15$ and $\alpha=-2.35$.
The cyan dashed and blue solid curves show the~\citet{2023arXiv230406658H} best-fit Schechter function obtained from the spectroscopically confirmed candidates at 
$z=9-12$ and extrapolated to $z\sim 16$, respectively. The upper limit (blue arrow) is obtained from photometric estimates at $z \sim 16$,  while the lower limits (cyan arrows) represent the spectroscopic constraints by \citet{2023arXiv230406658H}. The 21-cm signal constraints predict a much higher number of galaxies than the extrapolation of \citet{2023arXiv230406658H}'s results from $z=9-12$ by approximately 1-3 orders of magnitude at the faint end, depending on the presence and intensity of the radio background. 
}
    \label{fig:lfuv}
\end{figure}

The cosmic microwave background (CMB) spectrum is predicted to show an absorption feature at frequencies below 150 MHz imprinted when the Universe was flooded with Ly$\alpha$ photons emitted from the very first stars and before it was reheated and reionized \citep{madau97,tozzi00}.
The depth and timing (frequency) of the global 21-cm signal carry a wealth of information about the nature of the first sources and the thermal state of the intergalactic medium (IGM), and can constrain the physics of the very early Universe. The Experiment to Detect the Global EoR Signature (\textrm{EDGES}) team has reported a controversial detection~\citep{bowman18} of a flattened absorption profile in the sky-averaged radio spectrum, centered at 78 MHz and with an anomalous amplitude of 0.5 K, placing the birth of the first astrophysical sources at $z\sim 20$. 

Such a deep absorption trough implies 
new exotic physics during cosmic dawn, such as some interaction between dark-matter and baryons \citep[see e.g.,][]{Barkana18Nature, Munoz18, Slatyer2018}, or an excess radio background \citep[e.g.,][]{Fraser_2018, Pospelov_2018, Feng_2018, Ewall_2018, Fialkov_2019, Ewall2020}. It has also been argued \citep[see e.g.,][]{Hills_2018, Bradley2019, Singh2019, Sims2020} that the EDGES  signal may not be of astrophysical origin, and the recent non-detection by the SARAS3 experiment \citep{2022NatAs...6..607S} confirms earlier concerns.

While both \textrm{JWST} results based on early NIRCam observations in this extreme redshift regime and the nature of a radio absorption signal are highly uncertain, it is of interest to discuss the implications of a bright UV radiation background at cosmic dawn for 21-cm cosmology. 
In this Letter, we attempt to answer the following question: {\it Can the young galaxies detected by \textrm{JWST} at the highest redshifts provide enough Ly$\alpha$ radiation to mix the hyperfine levels of neutral hydrogen and produce a global 21-cm signal at the redshifts, $z\sim 17$,
probed by the \textrm{EDGES} and SARAS3 experiments?}
We  stress that, as in \citet{madau18}, our analysis focuses on the timing of such signal and on the constraints imposed by the required Wouthuysen–Field coupling strength on the UV radiation background at first light, and in the presence of different levels of radio background~\citep{Feng_2018, Ewall_2018}. Our analysis does not attempt to explain or dispute the absorption trough claimed by EDGES. 

\section{UV luminosity density at high redshift}

Figure~\ref{fig:rho} shows 
estimates of the UV luminosity density, $\rho_{\rm UV}$, from \textrm{HST} \citep{oesch18,ishigaki18,bouwens15}, \textrm{JWST}/NIRCam\footnote{We omit results at $z > 15$ based on a single candidate galaxy that was recently spectroscopically confirmed to be at lower redshift \citep{naidu_schrodingers_2022,zavala_dusty_2023,arrabal_haro_spectroscopic_2023}.}~\citep{donnan23,harikane22,bouwens2022} and \textrm{JWST}/NIRSpec \citep{2023arXiv230406658H},
and quoted in the legend for different magnitude faint-end cutoffs of $M_{\rm UV} = -18$ (left panel) and $M_{\rm UV} = -13$ (right panel)\footnote{We extrapolate only down to $M_{\rm UV} = -13$ since most of the reconstructed UV luminosity functions (LFs) from measurements and theory do not show  a cutoff at magnitudes brighter than this limit~\citep[e.g. see][]{2022ApJ...940...55B}.}.  All
values were obtained by integrating the observed LF down to the cutoff. Since the faint-end slope $\alpha$ at these early epochs is highly unconstrained, and these measured UV LFs are obtained at fixed $\alpha$, we assume a level of uncertainty inspired by~\citet{2022ApJ...940...55B}, where errors in $\alpha$ are shown to evolve from $\sim$ 1-2\%  at $z\sim 2-3$ to $\sim$ 4-7\% at $z \sim 7-10$. We then conservatively assume a 5\% and 10\% error in the faint-end slope of the galaxy LF at $z < 10$ and $z>10$, respectively, and add these uncertainties to JWST data only. 

In the figure we also plot the UV luminosity density required to produce a 21-cm feature at $16 < z < 19$ in the `minimal coupling' regime \citep[red box]{madau18}. This constraint is imposed on the background Ly$\alpha$ flux by the Wouthuysen–Field mechanism that mixes the hyperfine levels of neutral hydrogen and is key to the detectability of a 21-cm radio signal from the epoch of first light \citep{madau97}. It yields $\rho_{\rm UV} = 10^{24.52} - 10^{25}\, (18/(1+z))^{1/2}~{\rm erg\, s^{-1}\, Mpc^{-3}\, Hz^{-1}}$ for a proportionality constant ($g=0.06$) that relates the number density of Ly$\alpha$ photons to the ionizing emissivity (see Eq. 10 and associated text in~\citealt{madau18} for more details). In~\citet{madau18}, a fitting function, $\log_{10} (\rho_{\rm UV}/{\rm ergs^{-1} Mpc^{-3} Hz^{-1}}) = (26.30\pm 0.12)+ (-0.130 \pm 0.018)(z - 6)$ for a magnitude cut-off $M_{\rm UV} = -13$, was provided to describe a gradual redshift evolution consistent with  4$\leq z \leq$ 9 deep HST observations as well as the minimal coupling 21-cm regime. Since no JWST data were included in \citet{madau18}'s functional form, we refit for all data including HST, JWST and Wouthuysen–Field mechanism (red box). The updated fitting function,   $\log_{10} (\rho_{\rm UV}/{\rm ergs^{-1} Mpc^{-3} Hz^{-1}}) = (26.31\pm 0.16)+ (-0.118 \pm 0.019)(z - 6)$, 
and associated  1-$\sigma$ uncertainty are shown in the figure with the red solid line and shaded band. 

In addition to the minimal coupling 21-cm regime, we consider the same constraints in the presence of different levels of an additional (beyond the CMB) radio background  of brightness temperature $T_{\rm rad}$. Since the brightness temperature of the 21-cm signal scales 
as 
\begin{equation}
\delta T_{21}\propto 1-{T_{\rm CMB}+T_{\rm rad}\over T_s}, 
\end{equation}
where $T_s$ is the hydrogen spin temperature, the amplitude of the absorption signal can be increased by increasing $T_{\rm rad}$, 
leading to a multiplicative boost in the canonical absorption signal by the factor $F_{\rm boost} \approx  1 + T_{\rm rad}/T_{\rm CMB}$ in the  limit $T_s\ll T_{\rm CMB}$. First, we shall consider a  radiation excess by early black hole accretion as proposed by~\citet{Ewall_2018}, where a boost factor of $F_{\rm boost}\approx 3$ (corresponding to the presence of 1\% of the present day black hole mass at $z\sim 17$)
was found to reproduce the amplitude of the EDGES detection. This increases the Ly$\alpha$ coupling constraints on $\rho_{\rm UV}$
by the same factor, as shown by the brown box. In this scenario,
the best-fit UV luminosity density, $\log_{10} (\rho_{\rm UV}/{\rm ergs^{-1} Mpc^{-3} Hz^{-1}}) = (26.22\pm 0.15)+ (-0.072 \pm 0.017)(z - 6)$, has a much shallower redshift evolution.
Second, we consider the strong radiation excess detected by the Absolute Radiometer for Cosmology, Astrophysics and Diffuse Emission (ARCADE 2), which is consistent with the CMB at high frequencies and substantially higher than the CMB at low frequencies~\citep{2011ApJ...734....5F}. Following the fitting function provided by~\citet{Feng_2018}, we find a boost factor of $F_{\rm boost}\approx 20$, leading to a corresponding 
increase in the coupling constraint on $\rho_{\rm UV}$ (black box in the figure). This enhanced UV luminosity density
is comparable to existing  estimates at $z\sim 4-8$, and a flat evolution of $\log_{10} (\rho_{\rm UV}/{\rm ergs^{-1} Mpc^{-3} Hz^{-1}}) = 26.06\pm 0.24$ would reproduce both low and high-z constraints in this case. This extreme case of a flat evolution in $\rho_{\rm UV}$ is unlikely while the Universe is evolving from $z=20$ to $z=4$. However, we show this extreme case to set an upper limit for the expected shallow redshift evolution in the presence of radiation excess. 


Regardless of the magnitude cutoff, we find a general consistency between the early and more recent measurements of the UV luminosity density by \textrm{HST} and \textrm{JWST}. At the limit $M_{\rm UV} = -18$ (left panel in Figure~\ref{fig:rho}), HST and JWST data indicate a rapid decline in $\rho_{\rm UV}$ towards early epochs consistent with the evolving $\rho_{\rm UV}$ expected in constant star formation efficiency models~\citep{2023arXiv230406658H}, but inconsistent with the constraints imposed by a possible 
21-cm signal centered at redshift $z\sim17$. 
%
Extrapolating to fainter magnitudes and integrating down to $M_{\rm UV} = -13$, we find instead that the measurements suggest a milder evolution in $\rho_{\rm UV}$. This suggests that the high redshift constraints by \textrm{JWST} in the redshift range of $z\sim 8-12$ and a 21-cm signal in the minimal coupling regime at $z\sim 17$ may all be consistent with an extrapolation of the declining galaxy UV luminosity density measured at $z\sim 4-10$ by \textrm{HST}.
An even shallower decline in $\rho_{\rm UV}$ is required in the 
presence of a radio background excess from black holes or as detected by ARCADE 2. While uncertainties are still too large to rule out any of these scenarios, Figure~\ref{fig:rho} illustrates the potential of future JWST observations in placing independent constraints on exotic astrophysics during the epoch of the first light.

\begin{table*}[ht]
    \centering
    \begin{tabular}{ | l | c | c | }
    \hline
     & $\log_{10} (\rho_{\rm UV}/{\rm erg\, s^{-1}\, Mpc^{-3}\, Hz^{-1}})$ & $\log_{10}\, \phi^{\star}$  \\ \hline
    Minimal 21-cm coupling \citep{madau18} & $24.60\pm 0.24$ & $-4.671^{+0.240}_{-0.246}$ \\ \hline
    Radiation excess from black holes \citep{Ewall_2018} & $25.24\pm 0.24$ & $-4.194^{+0.243}_{-0.244}$ \\ \hline
    Radiation excess by ARCADE 2 \citep{Feng_2018} & $26.06\pm 0.24$ & $-3.381^{+0.244}_{-0.247}$ \\ \hline
    \end{tabular}
    \caption{ Constraints on the Schechter function parameter $\phi^{\star}$ from the UV luminosity density needed to produce a 21-cm signal at $z=16$ at fixed $M^{\star}_{\rm UV}=-19$ and $\alpha=-2.35$ in the minimal coupling regime and in the presence of different levels of a radio background.}
    \label{tab:phi}
\end{table*}

\section{The UV luminosity function at redshift 16}

It may be useful at this stage to understand what overall normalization of the galaxy UV LF would be required to produce a 21-cm feature at $z=16$ in the presence of different radio background excesses.
Using the minimal coupling constraints 
of \citet{madau18}, and fixing the Schechter LF parameters $M^{\star}_{\rm UV}=-19$ and $\alpha=-2.35$ (from fits provided by~\citealt{2023arXiv230406658H}), we derive at $z=16$ the normalization  $\log_{10}\phi^{\star}=-4.671^{+0.240}_{-0.246}$. Repeating the same procedure for the enhanced $\rho_{\rm UV}$
associated with the early black holes \citep{Ewall_2018} and  ARCADE 2 \citep{Feng_2018} radio excesses, we obtain  $\log_{10}\phi^{\star}=-4.194^{+0.243}_{-0.244}$ and $\log_{10}\phi^{\star}=-3.381^{+0.244}_{-0.247}$, respectively.  A summary of these constraints is provided in Table~\ref{tab:phi}.

The best-fit LF in the minimal coupling regime, shown as the solid red line in Figure~\ref{fig:lfuv}, lies below the 
LF at $z \sim 12$ calculated by \citet{2023arXiv230406658H} from spectroscopically confirmed candidates, as well as the upper limit at $z\sim 16$ \citep[]{2023arXiv230406658H}. This fit has a higher normalization, however, 
when compared to an extrapolation to $z \sim 16$ of the Schechter function parameters provided in ~\citet{2023arXiv230406658H}.
%
The boosted LF of the black hole radiation excess scenario at $z\sim 16$ (brown line) has an approximately similar amplitude/slope to the $z\sim 12 $ \citet{2023arXiv230406658H}'s spectroscopically measured LF at the faint-end, but declines more rapidly at the bright-end (compare $M^{\star}_{\rm UV} =-19$ vs. $M^{\star}_{\rm UV} =-20.3$). This fit still lies below the $z \sim 16$ photometric upper limit. The black curve in Figure~\ref{fig:lfuv} represents our prediction for LF of the ARCADE 2 radiation excess scenario, which is approximately one order of magnitude higher than the measured LF at $z\sim 12$. Since the latter is constructed using only lower limits, such a significantly boosted LF at $z\sim 16$ cannot be entirely ruled out.   
Future deep JWST surveys are expected to better probe the $z>12$ Universe \citep{wilkins_first_2023} and may provide a definitive test of these predictions.

\begin{figure*}\centering
    \includegraphics[scale=0.6]{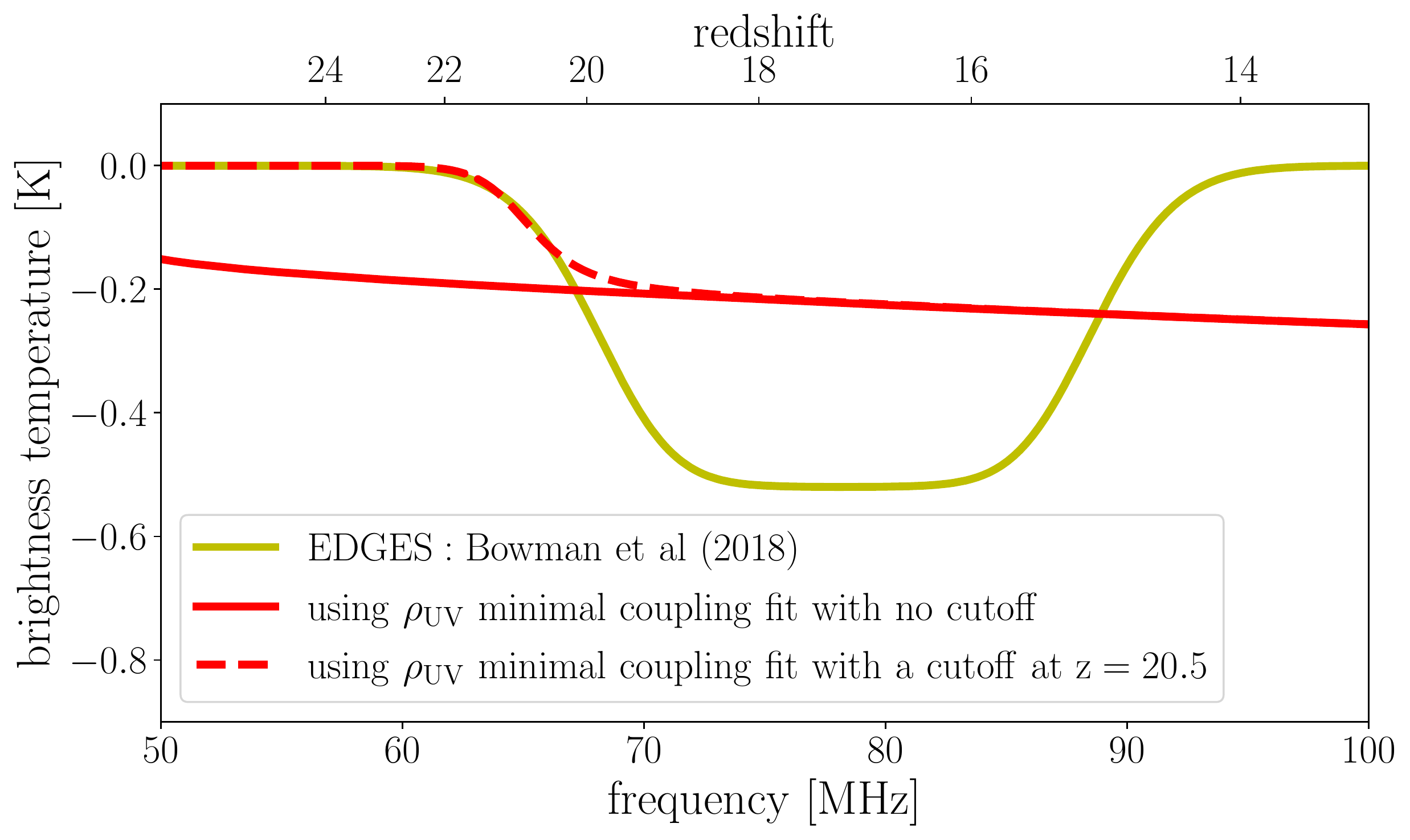}
\caption{Observed, sky-averaged brightness temperature at 21-cm. The red dashed and solid curves show the prediction from a minimal coupling scenario with a cut-off in the UV luminosity density at $z=20.5$ and without a cut-off, respectively.   
The yellow curve shows the spectral feature claimed by the EDGES experiment. The models ignore X-ray heating as well as the possible presence of an excess (over the CMB) radio background. 
}
    \label{fig:Tb}
\end{figure*}

\section{The 21-cm Signal}

In the previous section, we have shown how JWST data can constrain the presence of a 21-cm signal at extreme redshifts. Here, we offer a preliminary discussion of the few factors that may influence such feature. We use our default, minimal coupling scenario (see \citealt{madau18} and Fig. 1) for the evolution of the UV luminosity density to compute the expected 21-cm brightness temperature in the absence of X-ray heating and of a radio excess. 
Figure~\ref{fig:Tb} compares the EDGES claimed
absorption profile to predictions from a canonical model with a cut-off in the UV luminosity density at $z=20.5$ (dashed red curve) and without such cut-off (solid red curve). It shows how the absorption trough reported by the EDGES collaboration is several times stronger than that predicted by traditional astrophysical models, and the impact of a cut-off in $\rho_{\rm UV}$ on the onset of the global signal. As mentioned in the Introduction, the SARAS3~\citep{2022NatAs...6..607S} experiment 
has contradicted the EDGES detection. In the framework of our minimal coupling scenario, the SARAS non-detection may be explained by the shallower absorption signal predicted by the solid red curve in the figure, where the presence of Ly$\alpha$ sources at redshifts above 20 move the onset of 21-cm absorption to even lower frequencies. Alternatively, X-rays emission from the first generation of astrophysical sources may heat intergalactic gas above the temperature of the CMB, producing at these epochs a faint 21-cm signal in emission. Complications in assessing the impact of early X-ray heating on the detectability of 21-cm absorption include
the role of AGNs, the abundance of early X-ray binaries, the shape of the X-ray SED in the soft band \citep[e.g.,][]{Mesinger_2011, Fialkov_2014, Madau_2017}. We defer a detailed modeling of X-ray heating and radio excess scenarios to a future  paper.

\section{Conclusion}

The epoch of first light provides a unique window to the earliest astrophysical sources of radiation and their impact on the IGM. In this work, we have focused on the possibility that a high star formation efficiency at early times -- as implied by early \textrm{JWST} results -- may trigger the onset of intense Ly$\alpha$ emission at redshift $z=16-18$  and produce a cosmic 21-cm absorption signal 200 Myr after the Big Bang. We have shown that a radio signal at the frequencies probed by the EDGES  and SARAS experiment may be expected with an extrapolation of the evolving galaxy UV luminosity density measured at 4 $\lesssim z\lesssim 12$  by deep HST and JWST observations. If one integrates the UV LF  measured by  \textrm{JWST} down to $M_{\rm UV}=-13$, then all the observational data suggest a steady mild evolution of $\rho_{\rm UV}(z)$, generating at $z\sim 16-18$ enough Ly$\alpha$ photons to produce a global 21-cm signal via the Wouthuysen–Field effect. A milder evolution of $\rho_{\rm UV}(z)$, as required  by exotic models with a radio background excess over the CMB at early epochs may still be consistent with current JWST data given 
the large uncertainties associated with cosmic variance and the faint-end slope of the galaxy LF. Using a semi-analytical model of reionization,~\citet{2022arXiv220914312B} have recently shown that such a  mildly evolving luminosity density requires a much higher contribution from faint galaxies, since massive galaxies at high-$z$ are rare. 

We note that, using a fixed star formation efficiency linked to the halo mass function predicted by $\Lambda$CDM would lead to a significant drop in the UV luminosity density beyond $z > 12$, a decrease that is actually not observed \citep{sun_constraints_2016,harikane_goldrush_2018,2023arXiv230406658H,Mason_2018,mason_brightest_2023}. The high UV luminosity density inferred at  early times may require 
a revision of the standard astrophysics of early galaxy formation
\citep{lovell_extreme_2023,mason_brightest_2023,dekel_efficient_2023}, including the impact of dust \citep{ferrara_stunning_2022,nath_dust-free_2023}, a top heavy initial mass function or a high AGN fraction \citep{inayoshi_lower_2022,harikane_comprehensive_2022,2023arXiv230404348Y}, exotic sources such as primordial black holes or population III stars \citep{liu_accelerating_2022,wang_strong_2022,hutsi_did_2023,yuan_rapidly_2023,2022MNRAS.515.2901M}, or modifications to the cosmological model \citep{haslbauer_has_2022,menci_high-redshift_2022,maio_jwst_2023,biagetti_high-redshift_2023,dayal_warm_2023,melia_cosmic_2023}.

During the preparation of this Letter, \citet{2023arXiv230407085M} has independently 
discussed how the new \textrm{JWST} data may imply the presence of enough Ly$\alpha$ background photons to decouple the spin temperature from that of the CMB by redshift 14. In our work, we have focused on the redshift interval $z\sim 16-20$ where 21-cm experiments like \textrm{EDGES} and
SARAS are sensitive.


\section{Acknowledgements}
 The authors acknowledge the anonymous referee for the constructive feedback and suggestions that greatly improved the paper, and thank Adam Lidz, Martin Rey, Ingyin Zaw, Andrea Macci\`o, Anthony Pullen and Patrick Breysse for helpful discussions. 
SH acknowledges support for Program number HST-HF2-51507 provided by NASA through a grant from the Space Telescope Science Institute, which is operated by the Association of Universities for Research in Astronomy, incorporated, under NASA contract NAS5-26555.  CCL acknowledges support from a Dennis Sciama fellowship funded by the University of Portsmouth for the Institute of Cosmology and Gravitation.
PM acknowledges support from NASA TCAN grant 80NSSC21K0271 and the hospitality of New York University Abu Dhabi during the completion of this work. This research was supported in part by the National Science Foundation under Grant No. NSF PHY-1748958. Part of this work was completed during the KITP GALEVO23 workshop for data-driven astronomy.

%




\bibliography{sample631,jwst_edges_lovell}{}

\begin{thebibliography}{}
\expandafter\ifx\csname natexlab\endcsname\relax\def\natexlab#1{#1}\fi
\providecommand{\url}[1]{\href{#1}{#1}}
\providecommand{\dodoi}[1]{doi:~\href{http://doi.org/#1}{\nolinkurl{#1}}}
\providecommand{\doeprint}[1]{\href{http://ascl.net/#1}{\nolinkurl{http://ascl.net/#1}}}
\providecommand{\doarXiv}[1]{\href{https://arxiv.org/abs/#1}{\nolinkurl{https://arxiv.org/abs/#1}}}

\bibitem[{Adams {et~al.}(2022)Adams, Conselice, Ferreira, Austin, Trussler,
  Juodžbalis, Wilkins, Caruana, Dayal, Verma, \&
  Vijayan}]{adams_discovery_2022}
Adams, N.~J., Conselice, C.~J., Ferreira, L., {et~al.} 2022, arXiv:2207.11217,
  \dodoi{10.48550/arXiv.2207.11217}

\bibitem[{Arrabal~Haro {et~al.}(2023)Arrabal~Haro, Dickinson, Finkelstein,
  Kartaltepe, Donnan, Burgarella, Carnall, Cullen, Dunlop, Fernández,
  Fujimoto, Jung, Krips, Larson, Papovich, Pérez-González, Amorín, Bagley,
  Buat, Casey, Chworowsky, Cohen, Ferguson, Giavalisco, Huertas-Company,
  Hutchison, Kocevski, Koekemoer, Lucas, McLeod, McLure, Pirzkal, Trump,
  Weiner, Wilkins, \& Zavala}]{arrabal_haro_spectroscopic_2023}
Arrabal~Haro, P., Dickinson, M., Finkelstein, S.~L., {et~al.} 2023,
  Spectroscopic verification of very luminous galaxy candidates in the early
  universe, Tech. rep., \dodoi{10.48550/arXiv.2303.15431}

\bibitem[{Atek {et~al.}(2023)Atek, Shuntov, Furtak, Richard, Kneib, Mahler,
  Zitrin, McCracken, Charlot, Chevallard, \& Chemerynska}]{atek_revealing_2023}
Atek, H., Shuntov, M., Furtak, L.~J., {et~al.} 2023, MNRAS, 519, 1201,
  \dodoi{10.1093/mnras/stac3144}

\bibitem[{{Barkana}(2018)}]{Barkana18Nature}
{Barkana}, R. 2018, \nat, 555, 71, \dodoi{10.1038/nature25791}

\bibitem[{{Bera} {et~al.}(2022){Bera}, {Hassan}, {Smith}, {Cen}, {Garaldi},
  {Kannan}, \& {Vogelsberger}}]{2022arXiv220914312B}
{Bera}, A., {Hassan}, S., {Smith}, A., {et~al.} 2022, arXiv e-prints,
  arXiv:2209.14312, \dodoi{10.48550/arXiv.2209.14312}

\bibitem[{Biagetti {et~al.}(2023)Biagetti, Franciolini, \&
  Riotto}]{biagetti_high-redshift_2023}
Biagetti, M., Franciolini, G., \& Riotto, A. 2023, ApJ, 944, 113,
  \dodoi{10.3847/1538-4357/acb5ea}

\bibitem[{{Bouwens} {et~al.}(2022{\natexlab{a}}){Bouwens}, {Illingworth},
  {Ellis}, {Oesch}, \& {Stefanon}}]{2022ApJ...940...55B}
{Bouwens}, R.~J., {Illingworth}, G., {Ellis}, R.~S., {Oesch}, P., \&
  {Stefanon}, M. 2022{\natexlab{a}}, \apj, 940, 55,
  \dodoi{10.3847/1538-4357/ac86d1}

\bibitem[{{Bouwens} {et~al.}(2015){Bouwens}, {Illingworth}, {Oesch}, {Trenti},
  {Labb{\'e}}, {Bradley}, {Carollo}, {van Dokkum}, {Gonzalez}, {Holwerda},
  {Franx}, {Spitler}, {Smit}, \& {Magee}}]{bouwens15}
{Bouwens}, R.~J., {Illingworth}, G.~D., {Oesch}, P.~A., {et~al.} 2015, \apj,
  803, 34, \dodoi{10.1088/0004-637X/803/1/34}

\bibitem[{{Bouwens} {et~al.}(2022{\natexlab{b}}){Bouwens}, {Stefanon},
  {Brammer}, {Oesch}, {Herard-Demanche}, {Illingworth}, {Matthee}, {Naidu},
  {van Dokkum}, \& {van Leeuwen}}]{bouwens2022}
{Bouwens}, R.~J., {Stefanon}, M., {Brammer}, G., {et~al.} 2022{\natexlab{b}},
  arXiv e-prints, arXiv:2211.02607, \dodoi{10.48550/arXiv.2211.02607}

\bibitem[{{Bowman} {et~al.}(2018){Bowman}, {Rogers}, {Monsalve}, {Mozdzen}, \&
  {Mahesh}}]{bowman18}
{Bowman}, J.~D., {Rogers}, A. E.~E., {Monsalve}, R.~A., {Mozdzen}, T.~J., \&
  {Mahesh}, N. 2018, \nat, 555, 67, \dodoi{10.1038/nature25792}

\bibitem[{Boylan-Kolchin(2022)}]{boylan-kolchin_stress_2022}
Boylan-Kolchin, M. 2022, arXiv:2208.01611, \dodoi{10.48550/arXiv.2208.01611}

\bibitem[{Bradley {et~al.}(2022)Bradley, Coe, Brammer, Furtak, Larson,
  Andrade-Santos, Bhatawdekar, Bradac, Broadhurst, Carnall, Conselice, Diego,
  Frye, Fujimoto, Y.~Y~Hsiao, Hutchison, Jung, Mahler, McCandliss, Oguri,
  Postman, Sharon, Trenti, Vanzella, Welch, Windhorst, \&
  Zitrin}]{bradley_high-redshift_2022}
Bradley, L.~D., Coe, D., Brammer, G., {et~al.} 2022, High-{Redshift} {Galaxy}
  {Candidates} at \$z = 9-13\$ as {Revealed} by {JWST} {Observations} of
  {WHL0137}-08, Tech. rep., \dodoi{10.48550/arXiv.2210.01777}

\bibitem[{{Bradley} {et~al.}(2019){Bradley}, {Tauscher}, {Rapetti}, \&
  {Burns}}]{Bradley2019}
{Bradley}, R.~F., {Tauscher}, K., {Rapetti}, D., \& {Burns}, J.~O. 2019, \apj,
  874, 153, \dodoi{10.3847/1538-4357/ab0d8b}

\bibitem[{Castellano {et~al.}(2022{\natexlab{a}})Castellano, Fontana, Treu,
  Santini, Merlin, Leethochawalit, Trenti, Vanzella, Mestric, Bonchi, Belfiori,
  Nonino, Paris, Polenta, Roberts-Borsani, Boyett, Bradač, Calabrò,
  Glazebrook, Grillo, Mascia, Mason, Mercurio, Morishita, Nanayakkara,
  Pentericci, Rosati, Vulcani, Wang, \& Yang}]{castellano_early_2022}
Castellano, M., Fontana, A., Treu, T., {et~al.} 2022{\natexlab{a}}, The
  Astrophysical Journal, 938, L15, \dodoi{10.3847/2041-8213/ac94d0}

\bibitem[{Castellano {et~al.}(2022{\natexlab{b}})Castellano, Fontana, Treu,
  Merlin, Santini, Bergamini, Grillo, Rosati, Acebron, Leethochawalit, Paris,
  Bonchi, Belfiori, Calabrò, Nonino, Polenta, Trenti, Boyett, Broadhurst,
  Chen, Filippenko, Glazebrook, Mascia, Mason, Meneghetti, Mercurio, Metha,
  Morishita, Nanayakkara, Pentericci, Roberts-Borsani, Roy, Vanzella, Vulcani,
  Yang, \& Wang}]{castellano_high_2022}
---. 2022{\natexlab{b}}, A {High} {Density} of {Bright} {Galaxies} at
  \$z{\textbackslash}approx10\$ in the {A2744} region, Tech. rep.
\newblock \url{https://ui.adsabs.harvard.edu/abs/2022arXiv221206666C}

\bibitem[{Dayal \& Giri(2023)}]{dayal_warm_2023}
Dayal, P., \& Giri, S.~K. 2023, Warm dark matter constraints from the {JWST},
  Tech. rep., \dodoi{10.48550/arXiv.2303.14239}

\bibitem[{Dekel {et~al.}(2023)Dekel, Sarkar, Birnboim, Mandelker, \&
  Li}]{dekel_efficient_2023}
Dekel, A., Sarkar, K.~S., Birnboim, Y., Mandelker, N., \& Li, Z. 2023,
  Efficient {Formation} of {Massive} {Galaxies} at {Cosmic} {Dawn} by
  {Feedback}-{Free} {Starbursts}, Tech. rep., \dodoi{10.48550/arXiv.2303.04827}

\bibitem[{Donnan {et~al.}(2023)Donnan, McLeod, McLure, Dunlop, Carnall, Cullen,
  \& Magee}]{donnan_abundance_2023}
Donnan, C.~T., McLeod, D.~J., McLure, R.~J., {et~al.} 2023, MNRAS, 520, 4554,
  \dodoi{10.1093/mnras/stad471}

\bibitem[{Donnan {et~al.}(2022)Donnan, McLeod, Dunlop, McLure, Carnall, Begley,
  Cullen, Hamadouche, Bowler, McCracken, Milvang-Jensen, Moneti, \&
  Targett}]{donnan_evolution_2022}
Donnan, C.~T., McLeod, D.~J., Dunlop, J.~S., {et~al.} 2022, arXiv:2207.12356,
  \dodoi{10.48550/arXiv.2207.12356}

\bibitem[{{Donnan} {et~al.}(2023){Donnan}, {McLeod}, {Dunlop}, {McLure},
  {Carnall}, {Begley}, {Cullen}, {Hamadouche}, {Bowler}, {Magee}, {McCracken},
  {Milvang-Jensen}, {Moneti}, \& {Targett}}]{donnan23}
{Donnan}, C.~T., {McLeod}, D.~J., {Dunlop}, J.~S., {et~al.} 2023, \mnras, 518,
  6011, \dodoi{10.1093/mnras/stac3472}

\bibitem[{{Ewall-Wice} {et~al.}(2018){Ewall-Wice}, {Chang}, {Lazio},
  {Dor{\'e}}, {Seiffert}, \& {Monsalve}}]{Ewall_2018}
{Ewall-Wice}, A., {Chang}, T.~C., {Lazio}, J., {et~al.} 2018, \apj, 868, 63,
  \dodoi{10.3847/1538-4357/aae51d}

\bibitem[{{Ewall-Wice} {et~al.}(2020){Ewall-Wice}, {Chang}, \&
  {Lazio}}]{Ewall2020}
{Ewall-Wice}, A., {Chang}, T.-C., \& {Lazio}, T. J.~W. 2020, \mnras, 492, 6086,
  \dodoi{10.1093/mnras/stz3501}

\bibitem[{{Feng} \& {Holder}(2018)}]{Feng_2018}
{Feng}, C., \& {Holder}, G. 2018, \apjl, 858, L17,
  \dodoi{10.3847/2041-8213/aac0fe}

\bibitem[{Ferrara {et~al.}(2022)Ferrara, Pallottini, \&
  Dayal}]{ferrara_stunning_2022}
Ferrara, A., Pallottini, A., \& Dayal, P. 2022, On the stunning abundance of
  super-early, massive galaxies revealed by {JWST}, Tech. rep.
\newblock \url{https://ui.adsabs.harvard.edu/abs/2022arXiv220800720F}

\bibitem[{{Fialkov} \& {Barkana}(2019)}]{Fialkov_2019}
{Fialkov}, A., \& {Barkana}, R. 2019, \mnras, 486, 1763,
  \dodoi{10.1093/mnras/stz873}

\bibitem[{{Fialkov} {et~al.}(2014){Fialkov}, {Barkana}, \&
  {Visbal}}]{Fialkov_2014}
{Fialkov}, A., {Barkana}, R., \& {Visbal}, E. 2014, \nat, 506, 197,
  \dodoi{10.1038/nature12999}

\bibitem[{Finkelstein {et~al.}(2022)Finkelstein, Bagley, Haro, Dickinson,
  Ferguson, Kartaltepe, Papovich, Burgarella, Kocevski, Huertas-Company, Iyer,
  Koekemoer, Larson, Pérez-González, Rose, Tacchella, Wilkins, Chworowsky,
  Medrano, Morales, Somerville, Yung, Fontana, Giavalisco, Grazian, Grogin,
  Kewley, Kirkpatrick, Kurczynski, Lotz, Pentericci, Pirzkal, Ravindranath,
  Ryan, Trump, Yang, Team, Almaini, Amorín, Annunziatella, Backhaus, Barro,
  Behroozi, Bell, Bhatawdekar, Bisigello, Bromm, Buat, Buitrago, Calabrò,
  Casey, Castellano, Ortiz, Ciesla, Cleri, Cohen, Cole, Cooke, Cooper, Cooray,
  Costantin, Cox, Croton, Daddi, Davé, Vega, Dekel, Elbaz, Estrada-Carpenter,
  Faber, Fernández, Finkelstein, Freundlich, Fujimoto, García-Argumánez,
  Gardner, Gawiser, Gómez-Guijarro, Guo, Hamblin, Hamilton, Hathi, Holwerda,
  Hirschmann, Hutchison, Jaskot, Jha, Jogee, Juneau, Jung, Kassin, Bail, Leung,
  Lucas, Magnelli, Mantha, Matharu, McGrath, McIntosh, Merlin, Mobasher,
  Newman, Nicholls, Pandya, Rafelski, Ronayne, Santini, Seillé, Shah, Shen,
  Simons, Snyder, Stanway, Straughn, Teplitz, Vanderhoof, Vega-Ferrero, Wang,
  Weiner, Willmer, Wuyts, \& Zavala}]{finkelstein_long_2022}
Finkelstein, S.~L., Bagley, M.~B., Haro, P.~A., {et~al.} 2022, ApJL, 940, L55,
  \dodoi{10.3847/2041-8213/ac966e}

\bibitem[{Finkelstein {et~al.}(2023)Finkelstein, Bagley, Ferguson, Wilkins,
  Kartaltepe, Papovich, Yung, Haro, Behroozi, Dickinson, Kocevski, Koekemoer,
  Larson, Bail, Morales, Pérez-González, Burgarella, Davé, Hirschmann,
  Somerville, Wuyts, Bromm, Casey, Fontana, Fujimoto, Gardner, Giavalisco,
  Grazian, Grogin, Hathi, Hutchison, Jha, Jogee, Kewley, Kirkpatrick, Long,
  Lotz, Pentericci, Pierel, Pirzkal, Ravindranath, Ryan, Trump, Yang,
  Bhatawdekar, Bisigello, Buat, Calabrò, Castellano, Cleri, Cooper, Croton,
  Daddi, Dekel, Elbaz, Franco, Gawiser, Holwerda, Huertas-Company, Jaskot,
  Leung, Lucas, Mobasher, Pandya, Tacchella, Weiner, \&
  Zavala}]{finkelstein_ceers_2023}
Finkelstein, S.~L., Bagley, M.~B., Ferguson, H.~C., {et~al.} 2023, ApJL, 946,
  L13, \dodoi{10.3847/2041-8213/acade4}

\bibitem[{{Fixsen} {et~al.}(2011){Fixsen}, {Kogut}, {Levin}, {Limon}, {Lubin},
  {Mirel}, {Seiffert}, {Singal}, {Wollack}, {Villela}, \&
  {Wuensche}}]{2011ApJ...734....5F}
{Fixsen}, D.~J., {Kogut}, A., {Levin}, S., {et~al.} 2011, \apj, 734, 5,
  \dodoi{10.1088/0004-637X/734/1/5}

\bibitem[{{Fraser} {et~al.}(2018){Fraser}, {Hektor}, {H{\"u}tsi}, {Kannike},
  {Marzo}, {Marzola}, {Racioppi}, {Raidal}, {Spethmann}, {Vaskonen}, \&
  {Veerm{\"a}e}}]{Fraser_2018}
{Fraser}, S., {Hektor}, A., {H{\"u}tsi}, G., {et~al.} 2018, Physics Letters B,
  785, 159, \dodoi{10.1016/j.physletb.2018.08.035}

\bibitem[{{Harikane} {et~al.}(2023){Harikane}, {Nakajima}, {Ouchi}, {Umeda},
  {Isobe}, {Ono}, {Xu}, \& {Zhang}}]{2023arXiv230406658H}
{Harikane}, Y., {Nakajima}, K., {Ouchi}, M., {et~al.} 2023, arXiv e-prints,
  arXiv:2304.06658, \dodoi{10.48550/arXiv.2304.06658}

\bibitem[{Harikane {et~al.}(2018)Harikane, Ouchi, Ono, Saito, Behroozi, More,
  Shimasaku, Toshikawa, Lin, Akiyama, Coupon, Komiyama, Konno, Lin, Miyazaki,
  Nishizawa, Shibuya, \& Silverman}]{harikane_goldrush_2018}
Harikane, Y., Ouchi, M., Ono, Y., {et~al.} 2018, PASJ, 70, S11,
  \dodoi{10.1093/pasj/psx097}

\bibitem[{{Harikane} {et~al.}(2022){Harikane}, {Inoue}, {Mawatari},
  {Hashimoto}, {Yamanaka}, {Fudamoto}, {Matsuo}, {Tamura}, {Dayal}, {Yung},
  {Hutter}, {Pacucci}, {Sugahara}, \& {Koekemoer}}]{harikane22}
{Harikane}, Y., {Inoue}, A.~K., {Mawatari}, K., {et~al.} 2022, \apj, 929, 1,
  \dodoi{10.3847/1538-4357/ac53a9}

\bibitem[{Harikane {et~al.}(2022)Harikane, Ouchi, Oguri, Ono, Nakajima, Isobe,
  Umeda, Mawatari, \& Zhang}]{harikane_comprehensive_2022}
Harikane, Y., Ouchi, M., Oguri, M., {et~al.} 2022, arXiv:2208.01612,
  \dodoi{10.48550/arXiv.2208.01612}

\bibitem[{Haslbauer {et~al.}(2022)Haslbauer, Kroupa, Zonoozi, \&
  Haghi}]{haslbauer_has_2022}
Haslbauer, M., Kroupa, P., Zonoozi, A.~H., \& Haghi, H. 2022, ApJ, 939, L31,
  \dodoi{10.3847/2041-8213/ac9a50}

\bibitem[{{Hills} {et~al.}(2018){Hills}, {Kulkarni}, {Meerburg}, \&
  {Puchwein}}]{Hills_2018}
{Hills}, R., {Kulkarni}, G., {Meerburg}, P.~D., \& {Puchwein}, E. 2018, \nat,
  564, E32, \dodoi{10.1038/s41586-018-0796-5}

\bibitem[{Hütsi {et~al.}(2023)Hütsi, Raidal, Urrutia, Vaskonen, \&
  Veermäe}]{hutsi_did_2023}
Hütsi, G., Raidal, M., Urrutia, J., Vaskonen, V., \& Veermäe, H. 2023, Phys.
  Rev. D, 107, 043502, \dodoi{10.1103/PhysRevD.107.043502}

\bibitem[{Inayoshi {et~al.}(2022)Inayoshi, Harikane, Inoue, Li, \&
  Ho}]{inayoshi_lower_2022}
Inayoshi, K., Harikane, Y., Inoue, A.~K., Li, W., \& Ho, L.~C. 2022, ApJ, 938,
  L10, \dodoi{10.3847/2041-8213/ac9310}

\bibitem[{{Ishigaki} {et~al.}(2018){Ishigaki}, {Kawamata}, {Ouchi}, {Oguri},
  {Shimasaku}, \& {Ono}}]{ishigaki18}
{Ishigaki}, M., {Kawamata}, R., {Ouchi}, M., {et~al.} 2018, \apj, 854, 73,
  \dodoi{10.3847/1538-4357/aaa544}

\bibitem[{Labbe {et~al.}(2022)Labbe, van Dokkum, Nelson, Bezanson, Suess, Leja,
  Brammer, Whitaker, Mathews, \& Stefanon}]{labbe_very_2022}
Labbe, I., van Dokkum, P., Nelson, E., {et~al.} 2022, arXiv:2207.12446,
  \dodoi{10.48550/arXiv.2207.12446}

\bibitem[{Liu \& Bromm(2022)}]{liu_accelerating_2022}
Liu, B., \& Bromm, V. 2022, ApJ, 937, L30, \dodoi{10.3847/2041-8213/ac927f}

\bibitem[{Lovell {et~al.}(2023)Lovell, Harrison, Harikane, Tacchella, \&
  Wilkins}]{lovell_extreme_2023}
Lovell, C.~C., Harrison, I., Harikane, Y., Tacchella, S., \& Wilkins, S.~M.
  2023, MNRAS, 518, 2511, \dodoi{10.1093/mnras/stac3224}

\bibitem[{{Madau}(2018)}]{madau18}
{Madau}, P. 2018, \mnras, 480, L43, \dodoi{10.1093/mnrasl/sly125}

\bibitem[{{Madau} \& {Fragos}(2017)}]{Madau_2017}
{Madau}, P., \& {Fragos}, T. 2017, \apj, 840, 39,
  \dodoi{10.3847/1538-4357/aa6af9}

\bibitem[{{Madau} {et~al.}(1997){Madau}, {Meiksin}, \& {Rees}}]{madau97}
{Madau}, P., {Meiksin}, A., \& {Rees}, M.~J. 1997, \apj, 475, 429,
  \dodoi{10.1086/303549}

\bibitem[{Maio \& Viel(2023)}]{maio_jwst_2023}
Maio, U., \& Viel, M. 2023, A\&A, 672, A71, \dodoi{10.1051/0004-6361/202345851}

\bibitem[{Mason {et~al.}(2023)Mason, Trenti, \& Treu}]{mason_brightest_2023}
Mason, C.~A., Trenti, M., \& Treu, T. 2023, MNRAS, 521, 497,
  \dodoi{10.1093/mnras/stad035}

\bibitem[{Mason {et~al.}(2018)Mason, Treu, Dijkstra, Mesinger, Trenti,
  Pentericci, de~Barros, \& Vanzella}]{Mason_2018}
Mason, C.~A., Treu, T., Dijkstra, M., {et~al.} 2018, The Astrophysical Journal,
  856, 2, \dodoi{10.3847/1538-4357/aab0a7}

\bibitem[{{McCaffrey} {et~al.}(2023){McCaffrey}, {Hardin}, {Wise}, \&
  {Regan}}]{2023arXiv230413755M}
{McCaffrey}, J., {Hardin}, S., {Wise}, J., \& {Regan}, J. 2023, arXiv e-prints,
  arXiv:2304.13755, \dodoi{10.48550/arXiv.2304.13755}

\bibitem[{{Meiksin}(2023)}]{2023arXiv230407085M}
{Meiksin}, A. 2023, arXiv e-prints, arXiv:2304.07085.
\newblock \doarXiv{2304.07085}

\bibitem[{Melia(2023)}]{melia_cosmic_2023}
Melia, F. 2023, MNRAS: Letters, 521, L85, \dodoi{10.1093/mnrasl/slad025}

\bibitem[{Menci {et~al.}(2022)Menci, Castellano, Santini, Merlin, Fontana, \&
  Shankar}]{menci_high-redshift_2022}
Menci, N., Castellano, M., Santini, P., {et~al.} 2022, ApJ, 938, L5,
  \dodoi{10.3847/2041-8213/ac96e9}

\bibitem[{{Mesinger} {et~al.}(2011){Mesinger}, {Furlanetto}, \&
  {Cen}}]{Mesinger_2011}
{Mesinger}, A., {Furlanetto}, S., \& {Cen}, R. 2011, \mnras, 411, 955,
  \dodoi{10.1111/j.1365-2966.2010.17731.x}

\bibitem[{{Mittal} \& {Kulkarni}(2022)}]{2022MNRAS.515.2901M}
{Mittal}, S., \& {Kulkarni}, G. 2022, \mnras, 515, 2901,
  \dodoi{10.1093/mnras/stac1961}

\bibitem[{Morishita \& Stiavelli(2023)}]{morishita_physical_2023}
Morishita, T., \& Stiavelli, M. 2023, ApJ, 946, L35,
  \dodoi{10.3847/2041-8213/acbf50}

\bibitem[{{Mu{\~n}oz} {et~al.}(2018){Mu{\~n}oz}, {Dvorkin}, \&
  {Loeb}}]{Munoz18}
{Mu{\~n}oz}, J.~B., {Dvorkin}, C., \& {Loeb}, A. 2018, \prl, 121, 121301,
  \dodoi{10.1103/PhysRevLett.121.121301}

\bibitem[{Naidu {et~al.}(2022{\natexlab{a}})Naidu, Oesch, van Dokkum, Nelson,
  Suess, Whitaker, Allen, Bezanson, Bouwens, Brammer, Conroy, Illingworth,
  Labbe, Leja, Leonova, Matthee, Price, Setton, Strait, Stefanon, Tacchella,
  Toft, Weaver, \& Weibel}]{naidu_two_2022}
Naidu, R.~P., Oesch, P.~A., van Dokkum, P., {et~al.} 2022{\natexlab{a}},
  arXiv:2207.09434, \dodoi{10.48550/arXiv.2207.09434}

\bibitem[{Naidu {et~al.}(2022{\natexlab{b}})Naidu, Oesch, Setton, Matthee,
  Conroy, Johnson, Weaver, Bouwens, Brammer, Dayal, Illingworth, Barrufet,
  Belli, Bezanson, Bose, Heintz, Leja, Leonova, Marques-Chaves, Stefanon, Toft,
  van~der Wel, van Dokkum, Weibel, \& Whitaker}]{naidu_schrodingers_2022}
Naidu, R.~P., Oesch, P.~A., Setton, D.~J., {et~al.} 2022{\natexlab{b}},
  arXiv:2208.02794, \dodoi{10.48550/arXiv.2208.02794}

\bibitem[{Nath {et~al.}(2023)Nath, Vasiliev, Drozdov, \&
  Shchekinov}]{nath_dust-free_2023}
Nath, B.~B., Vasiliev, E.~O., Drozdov, S.~A., \& Shchekinov, Y.~A. 2023, MNRAS,
  521, 662, \dodoi{10.1093/mnras/stad505}

\bibitem[{{Oesch} {et~al.}(2018){Oesch}, {Bouwens}, {Illingworth}, {Labb{\'e}},
  \& {Stefanon}}]{oesch18}
{Oesch}, P.~A., {Bouwens}, R.~J., {Illingworth}, G.~D., {Labb{\'e}}, I., \&
  {Stefanon}, M. 2018, \apj, 855, 105, \dodoi{10.3847/1538-4357/aab03f}

\bibitem[{{Pospelov} {et~al.}(2018){Pospelov}, {Pradler}, {Ruderman}, \&
  {Urbano}}]{Pospelov_2018}
{Pospelov}, M., {Pradler}, J., {Ruderman}, J.~T., \& {Urbano}, A. 2018, \prl,
  121, 031103, \dodoi{10.1103/PhysRevLett.121.031103}

\bibitem[{{Prada} {et~al.}(2023){Prada}, {Behroozi}, {Ishiyama}, {Klypin}, \&
  {P{\'e}rez}}]{2023arXiv230411911P}
{Prada}, F., {Behroozi}, P., {Ishiyama}, T., {Klypin}, A., \& {P{\'e}rez}, E.
  2023, arXiv e-prints, arXiv:2304.11911, \dodoi{10.48550/arXiv.2304.11911}

\bibitem[{Pérez-González {et~al.}(2023)Pérez-González, Costantin,
  Langeroodi, Rinaldi, Annunziatella, Ilbert, Colina, Noorgaard-Nielsen, Greve,
  Ostlin, Wright, Alonso-Herrero, Álvarez Márquez, Caputi, Eckart, Le~Fèvre,
  Labiano, García-Marín, Hjorth, Kendrew, Pye, Tikkanen, van~der Werf,
  Walter, Ward, Bosman, Gillman, García-Argumánez, \&
  María~Mérida}]{perez-gonzalez_life_2023}
Pérez-González, P.~G., Costantin, L., Langeroodi, D., {et~al.} 2023, Life
  beyond 30: probing the -20, Tech. rep.
\newblock \url{https://ui.adsabs.harvard.edu/abs/2023arXiv230202429P}

\bibitem[{Robertson(2022)}]{robertson_galaxy_2022}
Robertson, B.~E. 2022, ARAA, 60, 121,
  \dodoi{10.1146/annurev-astro-120221-044656}

\bibitem[{{Sims} \& {Pober}(2020)}]{Sims2020}
{Sims}, P.~H., \& {Pober}, J.~C. 2020, \mnras, 492, 22,
  \dodoi{10.1093/mnras/stz3388}

\bibitem[{{Singh} \& {Subrahmanyan}(2019)}]{Singh2019}
{Singh}, S., \& {Subrahmanyan}, R. 2019, \apj, 880, 26,
  \dodoi{10.3847/1538-4357/ab2879}

\bibitem[{{Singh} {et~al.}(2022){Singh}, {Jishnu}, {Subrahmanyan}, {Udaya
  Shankar}, {Girish}, {Raghunathan}, {Somashekar}, {Srivani}, \&
  {Sathyanarayana Rao}}]{2022NatAs...6..607S}
{Singh}, S., {Jishnu}, N.~T., {Subrahmanyan}, R., {et~al.} 2022, Nature
  Astronomy, 6, 607, \dodoi{10.1038/s41550-022-01610-5}

\bibitem[{{Slatyer} \& {Wu}(2018)}]{Slatyer2018}
{Slatyer}, T.~R., \& {Wu}, C.-L. 2018, \prd, 98, 023013,
  \dodoi{10.1103/PhysRevD.98.023013}

\bibitem[{Sun \& Furlanetto(2016)}]{sun_constraints_2016}
Sun, G., \& Furlanetto, S.~R. 2016, MNRAS, 460, 417,
  \dodoi{10.1093/mnras/stw980}

\bibitem[{{Tozzi} {et~al.}(2000){Tozzi}, {Madau}, {Meiksin}, \&
  {Rees}}]{tozzi00}
{Tozzi}, P., {Madau}, P., {Meiksin}, A., \& {Rees}, M.~J. 2000, \apj, 528, 597,
  \dodoi{10.1086/308196}

\bibitem[{Wang {et~al.}(2022)Wang, Cheng, Ge, Meng, Daddi, Yan, Jones, Malkan,
  Arrabal~Haro, Brammer, \& Oguri}]{wang_strong_2022}
Wang, X., Cheng, C., Ge, J., {et~al.} 2022, A strong {He} {II}
  \${\textbackslash}lambda\$1640 emitter with extremely blue {UV} spectral
  slope at \$z=8.16\$: presence of {Pop} {III} stars?, Tech. rep.
\newblock \url{https://ui.adsabs.harvard.edu/abs/2022arXiv221204476W}

\bibitem[{Wilkins {et~al.}(2023)Wilkins, Vijayan, Lovell, Roper, Irodotou,
  Caruana, Seeyave, Kuusisto, Thomas, \& Parris}]{wilkins_first_2023}
Wilkins, S.~M., Vijayan, A.~P., Lovell, C.~C., {et~al.} 2023, MNRAS, 519, 3118,
  \dodoi{10.1093/mnras/stac3280}

\bibitem[{Yan {et~al.}(2023)Yan, Ma, Ling, Cheng, \& Huang}]{yan_first_2023}
Yan, H., Ma, Z., Ling, C., Cheng, C., \& Huang, J.-S. 2023, ApJ, 942, L9,
  \dodoi{10.3847/2041-8213/aca80c}

\bibitem[{Yuan {et~al.}(2023)Yuan, Lei, Wang, Wang, Wang, Chen, Shen, Cai, \&
  Fan}]{yuan_rapidly_2023}
Yuan, G.-W., Lei, L., Wang, Y.-Z., {et~al.} 2023, Rapidly growing primordial
  black holes as seeds of the massive high-redshift {JWST} {Galaxies},  arXiv,
  \dodoi{10.48550/arXiv.2303.09391}

\bibitem[{{Yung} {et~al.}(2023){Yung}, {Somerville}, {Finkelstein}, {Wilkins},
  \& {Gardner}}]{2023arXiv230404348Y}
{Yung}, L.~Y.~A., {Somerville}, R.~S., {Finkelstein}, S.~L., {Wilkins}, S.~M.,
  \& {Gardner}, J.~P. 2023, arXiv e-prints, arXiv:2304.04348,
  \dodoi{10.48550/arXiv.2304.04348}

\bibitem[{Zavala {et~al.}(2023)Zavala, Buat, Casey, Finkelstein, Burgarella,
  Bagley, Ciesla, Daddi, Dickinson, Ferguson, Franco, Jiménez-Andrade,
  Kartaltepe, Koekemoer, Bail, Murphy, Papovich, Tacchella, Wilkins, Aretxaga,
  Behroozi, Champagne, Fontana, Giavalisco, Grazian, Grogin, Kewley, Kocevski,
  Kirkpatrick, Lotz, Pentericci, Pérez-González, Pirzkal, Ravindranath,
  Somerville, Trump, Yang, Yung, Almaini, Amorín, Annunziatella, Haro,
  Backhaus, Barro, Bell, Bhatawdekar, Bisigello, Buitrago, Calabrò,
  Castellano, Chávez~Ortiz, Chworowsky, Cleri, Cohen, Cole, Cooke, Cooper,
  Cooray, Costantin, Cox, Croton, Davé, de~La~Vega, Dekel, Elbaz,
  Estrada-Carpenter, Fernández, Finkelstein, Freundlich, Fujimoto,
  García-Argumánez, Gardner, Gawiser, Gómez-Guijarro, Guo, Hamilton, Hathi,
  Holwerda, Hirschmann, Huertas-Company, Hutchison, Iyer, Jaskot, Jha, Jogee,
  Juneau, Jung, Kassin, Kurczynski, Larson, Leung, Long, Lucas, Magnelli,
  Mantha, Matharu, McGrath, McIntosh, Medrano, Merlin, Mobasher, Morales,
  Newman, Nicholls, Pandya, Rafelski, Ronayne, Rose, Ryan, Santini, Seillé,
  Shah, Shen, Simons, Snyder, Stanway, Straughn, Teplitz, Vanderhoof,
  Vega-Ferrero, Wang, Weiner, Willmer, Wuyts, \& {(The Ceers
  Team)}}]{zavala_dusty_2023}
Zavala, J.~A., Buat, V., Casey, C.~M., {et~al.} 2023, ApJ, 943, L9,
  \dodoi{10.3847/2041-8213/acacfe}

\end{thebibliography}
\bibliographystyle{aasjournal}



\end{document}